\documentclass[12pt]{article}

\pdfoutput=1

\usepackage{cite}

\usepackage{amssymb,latexsym}

\usepackage{epstopdf}

\usepackage{graphics,epsfig}

\usepackage{color}

\usepackage{mathrsfs}

\DeclareMathAlphabet{\mathpzc}{OT1}{pzc}{m}{it}

\let\a=\alpha \let\b=\beta \let\g=\gamma \let\d=\delta \let\e=\epsilon
\let\z=\zeta  \let\th=\theta  \let\k=\kappa
\let\l=\lambda \let\m=\mu \let\n=\nu \let\x=\xi \let\p=\pi 
\let\s=\sigma   \let\f=\phi  
 
\let\w=\omega      \let\G=\Gamma  \let\Th=\Theta 
\let\X=\Xi  \let\S=\Sigma  \let\Y=\Psi
 
\let\la=\label  
  
\def\nn{\nonumber} \def\bd{\begin{document}} \def\ed{\end{document}}
\def\ds{\documentstyle} \let\fr=\frac \let\bl=\bigl \let\br=\bigr
\let\Br=\Bigr \let\Bl=\Bigl
\let\bm=\bibitem
\let\na=\nabla
\def\tU{{\widetilde U}}
\let\pa=\partial \let\ov=\overline
\def\ie{{\it i.e.\ }}
\newcommand{\be}{\begin{equation}}
\newcommand{\ee}{\end{equation}}
\def\ba{\begin{array}}
\def\ea{\end{array}}
\def\ft#1#2{{\textstyle{{\scriptstyle #1}\over {\scriptstyle #2}}}}
\def\fft#1#2{{#1 \over #2}}
\def\F#1#2{{ F_{#1}^{(#2)} }}
\def\cF#1#2{{ {\cal F}_{#1}^{(#2)} }}

\def\R{{\bf R}}
\def\sst#1{{\scriptscriptstyle #1}}
\def\oneone{\rlap 1\mkern4mu{\rm l}}
\def\e7{E_{7(+7)}}
\def\td{\tilde}
\def\wtd{\widetilde}
\def\im{{\rm i}}
\def\bog{Bogomol'nyi\ }
\newcommand{\ho}[1]{$\, ^{#1}$}
\newcommand{\hoch}[1]{$\, ^{#1}$}
\newcommand{\bea}{\begin{eqnarray}}
\newcommand{\eea}{\end{eqnarray}}
\newcommand{\ra}{\rightarrow}
\newcommand{\lra}{\longrightarrow}
\newcommand{\Lra}{\Leftrightarrow}
\newcommand{\ap}{\alpha^\prime}
\newcommand{\bp}{\tilde \beta^\prime}
\newcommand{\cB}{{\cal B}}
\newcommand{\cO}{{\cal O}}
\newcommand{\vecx}{\vec{x}}
\newcommand{\vecy}{\vec{y}}
\newcommand{\vecp}{\vec{p}}
\newcommand{\vecq}{\vec{q}}
\newcommand{\tr}{{\rm tr} }
\newcommand{\Tr}{{\rm Tr} }
\newcommand{\NP}{Nucl. Phys. }

\newcommand{\cL}{{\cal L}}
\newcommand{\cA}{{\cal A}}
\newcommand{\cT}{{\cal T}}
\newcommand{\cD}{{\cal D}}
\newcommand{\cH}{{\cal H}}

\def\sst#1{{\scriptscriptstyle #1}}
\def\0{{\sst{(0)}}}
\def\1{{\sst{(1)}}}
\def\2{{\sst{(2)}}}
\def\3{{\sst{(3)}}}
\def\4{{\sst{(4)}}}
\def\5{{\sst{(5)}}}
\def\6{{\sst{(6)}}}
\def\7{{\sst{(7)}}}
\def\8{{\sst{(8)}}}
\def\9{{\sst{(9)}}}
\def\p{{\sst{(p)}}}
\def\q{{\sst{(q)}}}
\def\ve{\varepsilon}
\def\vf{\varphi}
\def\F{\Phi}
\def\wg{\wedge}

\def\thb{\bar{\theta}}
\def\Thb{\bar{\Theta}}
\def\barp{\bar{p}}
\def\barq{\bar{q}}
\def\barc{\bar{c}}
\def\bard{\bar{d}}
\def\e{\epsilon}

\def \bi{\bibitem}
\def \la {\label}

\def \l {\lambda}
\def\foot{\footnote}
\def \tl  {{\tilde \l}}
\def \sql {{\sqrt \l}}
\def \adss {$AdS_5 \times S^5$\ }
\newcommand{\rf}[1]{(\ref{#1})}
\def \ov {\over}

\def\th{\theta}
\def\Th{\Theta}
\def\vth{\vartheta}
\def\btheta{{\bar\theta}}
\def\ttheta{{{\tilde\theta}}}
\def\bttheta{{{\bar\ttheta}}}
\def\vth{\vartheta}

\def\ra{\rightarrow}
\def\N{\nabla}
\def\F{{\cal F}}
\def\uM{\underline{M}}
\def\uA{\underline{A}}
\def\uN{\underline{N}}
\def\uP{\underline{P}}
\def\ua{\underline{a}}
\def\ub{\underline{b}}
\def\uc{\underline{c}}
\def\ud{\underline{d}}
\def\ue{\underline{e}}
\def\uf{\underline{f}}
\def\ui{\underline{i}}
\def\uj{\underline{j}}
\def\uk{\underline{k}}
\def\ul{\underline{l}}
\def\ual{\underline{\alpha}}
\def\ube{\underline{\beta}}
\def\um{\underline{m}}
\def\un{\underline{n}}
\def\up{\underline{p}}
\def\uq{\underline{q}}
\def\ur{\underline{r}}
\def\us{\underline{s}}
\def\umu{\underline{\mu}}
\def\unu{\underline{\nu}}
\def\ula{\underline{\l}}
\def\uka{\underline{\k}}
\def\usi{\underline{\s}}
\def\urh{\underline{\r}}
\def\cc{\circ}
\def\eqv{\equiv}

\def\ni{\noindent}

\def\Ep{E^{{}^{(+)}}}
\def\Em{E^{{}^{(-)}}}

\def\Mp{M^{{}^{(+)}}}
\def\Mm{M^{{}^{(-)}}}

\def \ha{{1\ov 2}}

\def\r{\rho}

\def\Y{{\rm Y}}
\def\X{{\rm X}}
\def\tY{\tilde{\rm Y}}
\def\tX{\tilde{\rm X}}
\def\dY{\dot{\rm Y}}
\def\dX{\dot{\rm X}}

\def \J {\mathcal{J}}
\def \del {\partial}

\def\dF{\dot{F}}
\def\dG{\dot{G}}
\def\df{\dot{f}}
\def \E {{\cal E}}
\def \S {{\cal S}}
\def \J {{\cal J}}

\def\ms{\mathcal{S}}
\def\mj{\mathcal{J}}
\def\soj{\fr{\ms}{\mj}}
\def \R {{\bf R}}
\def \om {\omega}
\def \bE {\bar E}
\def \x {{\cal X}}

\def \bi{\bibitem}
\def \la {\label}

\def \l {\lambda}
\def\foot{\footnote}
\def \tl  {{\tilde \l}}
\def \sql {{\sqrt \l}}
\def \adss {$AdS_5 \times S^5$\ }
\def \ov {\over}

\def \varpi {{\rm w}}

\def\thb{\bar{\theta}}
\def\Thb{\bar{\Theta}}
\def\mb{\bar{\m}}
\def\ab{\bar{\a}}
\def\zb{\bar{z}}
\def\psib{\bar{\psi}}
\def\barp{\bar{p}}
\def\barq{\bar{q}}
\def\barc{\bar{c}}
\def\bard{\bar{d}}
\def\e{\epsilon}
\def\wb{\bar{w}}
\def\lb{\bar{\l}}
\def\Jb{\bar{J}}
\def\Nb{\bar{N}}
\def\Zb{\bar{Z}}
\def\pab{\bar{\pa}}

\def\At{\tilde{A}}
\def\Bt{\tilde{B}}
\def\Ct{\tilde{C}}
\def\Dt{\tilde{D}}
\def\Et{\tilde{E}}
\def\Ft{\tilde{F}}
\def\Gt{\tilde{G}}
\def\Ht{\tilde{H}}
\def\Kt{\tilde{K}}
\def\Mt{\tilde{M}}
\def\Nt{\tilde{N}}
\def\Rt{\tilde{R}}
\def\at{\tilde{a}}
\def\bt{\tilde{b}}
\def\ct{\tilde{c}}
\def\dt{\tilde{d}}
\def\et{\tilde{e}}
\def\ft{\tilde{f}}
\def\htil{\tilde{h}}
\def\gt{\tilde{g}}
\def\nt{\tilde{n}}
\def\mut{\tilde{\mu}}
\def\nut{\tilde{\nu}}
\def\pht{\tilde{\f}}
\def\vft{\tilde{\vf}}

\def\rht{\tilde{\rho}}

\def\asth{\hat{*}}
\def\phh{\hat{\phi}}

\def\bA{{\bf A}}

\def\ola{\overleftarrow}
\def\ora{\overrightarrow}
\def\alt{\tilde{\a}}

\def\eh{\hat{e}}
\def\eph{\hat{\e}}
\def\ph{\hat{p}}
\def\alh{\hat{\a}}
\def\beh{\hat{\b}}
\def\gah{\hat{\g}}
\def\Fh{\hat{F}}
\def\muh{\hat{\m}}
\def\nuh{\hat{\n}}
\def\thh{\hat{\th}}
\def\rhh{\hat{\r}}
\def\dh{\hat{d}}
\def\ih{\hat{i}}
\def\jh{\hat{j}}
\def\hh{\hat{h}}
\def\nh{\hat{n}}
\def\gh{\hat{g}}
\def\kh{\hat{k}}
\def\deh{\hat{\d}}
\def\wh{\hat{w}}
\def\lah{\hat{\l}}
\def\Ah{\hat{A}}
\def\Kh{\hat{K}}
\def\Nh{\hat{N}}
\def\Rh{\hat{R}}
\def\Ch{\hat{C}}
\def\Omh{\hat{\Omega}}

\def\xh{\hat{x}}

\def\ps{\rlap{\, /}\;\,p }
\def\ks{\rlap{\, /}\;\,k }

\def\gym{g_{YM}}

\def\adot{\dot{a}}
\def\bdot{\dot{b}}
\def\bpa{\bar{\pa}}

\def\pr{\prime}
\def\ssk{\medskip}
\def\clb{\color{blue}}
\def\clr{\color{red}}
\def\clg{\color{green}}

\def\bfA{{\bf A}}
\def\bfB{{\bf B}}
\def\bfK{{\bf K}}
\def\bfU{{\bf U}}
\def\bfX{{\bf X}}
\def\bfY{{\bf Y}}
\def\bfZ{{\bf Z}}
\def\bfg{{\bf g}}
\def\bfn{{\bf n}}

\begin{document}

\overfullrule=0pt
\parskip=2pt
\parindent=12pt
\headheight=0in \headsep=0in \topmargin=0in
\oddsidemargin=0in

\vspace{ -3cm}
\thispagestyle{empty}

 \vspace{0.1cm}

\setcounter{equation}{0}
\setcounter{footnote}{0}
\setcounter{section}{0}

\begin{center}

{\Large\bf Foliation, jet bundle and quantization of Einstein gravity}

\vskip 0.8cm

 \vspace{.5cm}

\vspace{0.5cm}
I. Y. Park
\\

\vspace{0.3cm}


{\it Department of Physics, Hanyang University \\
Seoul 133-791, Korea}\\

\vspace{0.3cm}
{\it Department of Applied Mathematics,
Philander Smith College 
                               \\
Little Rock, AR 72223, USA \\
inyongpark05@gmail.com
}

\end{center}

 \vspace{0.1cm}

\begin{abstract}

In \cite{Park:2014tia} we proposed a way of quantizing gravity with the Hamiltonian and Lagrangian analyses in the ADM setup.
One of the key observations was that the physical configuration space of 
the 4D Einstein-Hilbert action admits a three-dimensional description, {thereby making gravity renormalization possible through a metric field redefinition.} Subsequently, a more mathematical and complementary picture of the reduction based on foliation theory was presented  in \cite{Park:2014qoa}. With the setup of foliation the physical degrees of freedom have been identified with a certain leaf. Here we expand the work of \cite{Park:2014qoa} by adding another mathematical ingredient - an element of jet bundle theory. With the introduction of the jet bundle, the procedure of identifying the true degrees of freedom outlined therein is made precise and the whole picture of the reduction is put on firm mathematical ground.

\end{abstract}
\newpage

\section{Introduction}

There have been two main approaches in tackling the quantization of Einstein gravity; namely, canonical and covariant (see, e.g.,\cite{Carlip:2001wq,Woodard:2014jba,Thiemann:2007zz} for reviews\footnote{There are other approaches. Notably there is a lattice-type approach based on spacetime triangulation \cite{Ambjorn:2012jv}.}). At an early stage, the canonical approach was pursued within the Hamiltonian formulation, and led to the Wheeler-DeWitt equation.
Later, its main theme became the so-called configuration space reduction \cite{Cendra}\cite{Marsden} that employs the machinery of differential geometry, in particular, symplectic geometry and jet bundle. Meanwhile the covariant approach followed a path within a more conventional physics framework. The main endeavors along this line were enumeration of the counter terms in the effective action \cite{'tHooft:1974bx,Deser:1974cz,Goroff:1985th,Stelle:1976gc,Antoniadis:1986tu} and the progress made in the asymptotically safe gravity \cite{Weinberg3,Reuter:1996cp,Niedermaier:2006ns,Litim:2008tt,Percacci:2011fr}. It was established for 4D Einstein action that the divergences do not cancel except for one-loop; one faces proliferation of counter terms as the order of loop increases - which turns out to be typical of other gravity theories, and the theory loses its predictability.

A question has recently been raised \cite{Park:2014tia} regarding the conventional framework of the Feynman diagram computation in which the non-dynamical fields contribute, under the umbrella of the covariant approach, to loop diagrams and appear as external states as well. The number of degrees of freedom (i.e., the number of a metric components) of a 4D metric is ten to start. One gauge-fixes the 4D gauge symmetry thereby effectively removing four degrees of freedom. This means that six metric components run around the loop diagrams. In quantization and diagrammatic analysis, one first examines the physical states of the theory (usually) through the canonical Hamiltonian analysis. In the canonical Hamiltonian analysis carried out in the ADM setup (which had been introduced with the goal of separating out the dynamical degrees of freedom), it was revealed that some of the metric components, named the lapse function and shift vector, are non-dynamical, a result that is well known by now. Therefore only two out of ten components are dynamical. An approach in which all of the unphysical degrees of freedom are removed has been proposed 
with the observation  \cite{Park:2014tia,Park:2014qoa,Park:2014noa} that
the non-dynamism of the shift vector and lapse function leads to 
effective reduction of 4D gravity when expanded around relatively simple vacuua. (A more precise characterization of these vacuua will be discussed below.)

The approach of \cite{Park:2014tia,Park:2014qoa,Park:2014noa} has features of both the canonical and covariant approaches: it employs the 3+1 splitting as does the canonical approach and a fixed background is considered as in the covariant approach.
 The analysis in \cite{Park:2014tia} was carried out in the ADM formalism. 
After starting with the ADM Lagrangian, the (more or less standard) Dirac's Hamiltonian formulation was employed, and the Lagrangian setup was revisited afterwards. Based on the fact that the lapse function and shift vector are non-dynamical, it was proposed in \cite{Park:2014tia}\cite{Park:2014noa} that the shift vector be gauge-fixed by using the 3D residual symmetry that remains after the 4D bulk gauge-fixing through the de Donder gauge. The shift vector gauge-fixing introduces a constraint - analogous to the momentum constraint in the Hamiltonian quantization - in the ADM Lagrangian setup. (The method of \cite{Park:2014tia} is applicable to a class of relatively simple backgrounds. A more precise characterization will be given in section 4 below.)   
The relevance of foliation theory (reviews on foliation theory can be found, e.g., in \cite{Molino,Moerdijk,Gromoll,Rovenskii,Candel,Dotto}) was recognized while examining possible implications of the shift vector constraint; it was realized that the shift vector constraint implies that the foliation of the spacetime should be of a special type known as Riemannian in foliation theory.
Interestingly, Riemannian foliation admits another special foliation, dual totally geodesic foliation, a result relatively recent in the timeframe of mathematics \cite{Cairns}. One of the facts that makes these special foliations interesting is the presence of the so-called parallelism \cite{Molino}, and in the context of the totally geodesic foliation under consideration the parallelism is ``tangential" \cite{Cairns} and has the associated abelian Lie algebra.

\begin{figure}
\centerline{
\begin{minipage}[b]{8cm}
             \epsfxsize=8cm
              \epsfbox{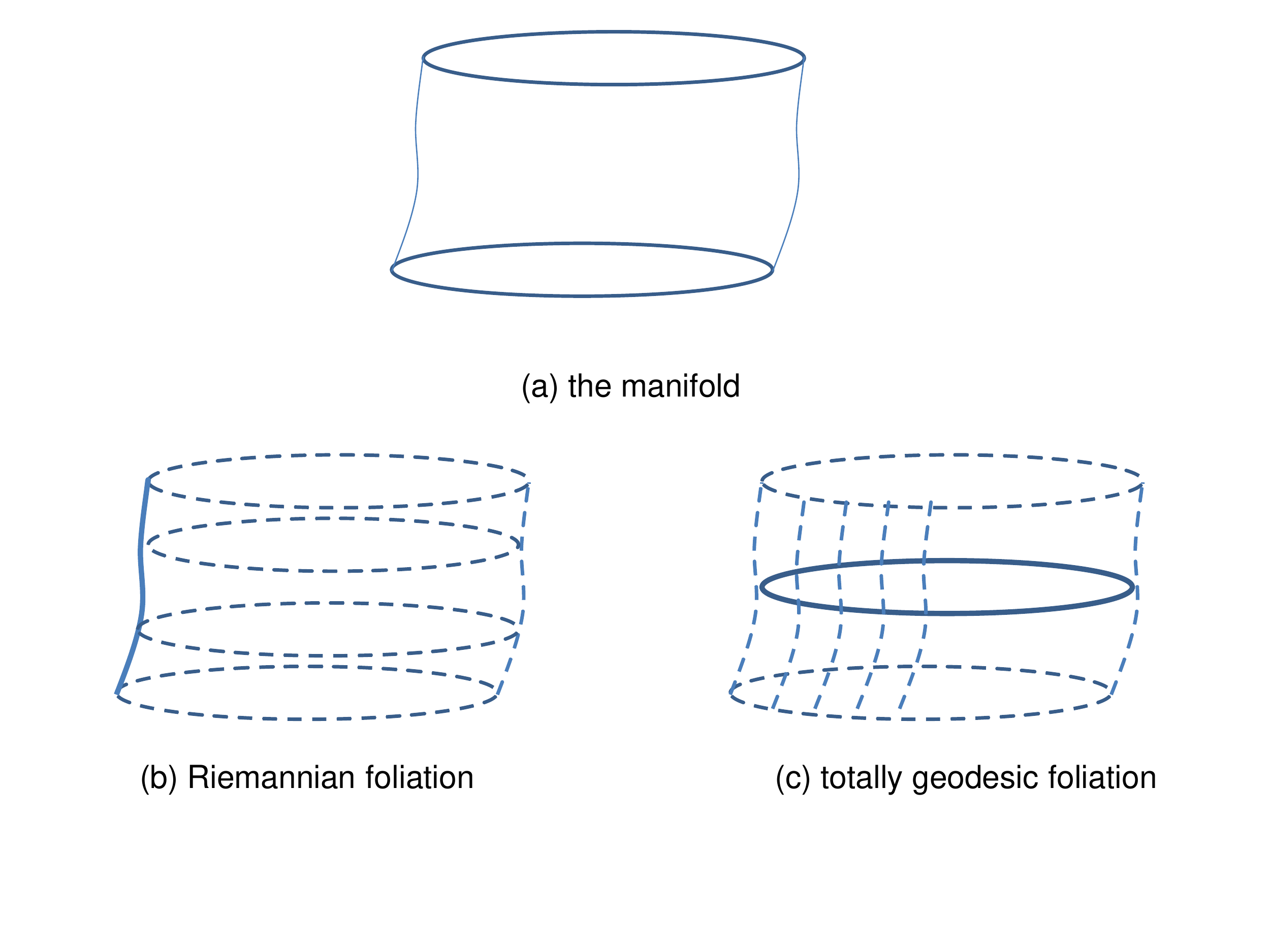}
      \end{minipage}
      }
\caption{duality in foliations: the solid lines in (b) and (c) represent the base (i.e., the space of the leaves) in each case}
\label{fig1}
\end{figure}

The proposal in \cite{Park:2014tia} has a complementary, more mathematical version \cite{Park:2014qoa} that relies crucially on this duality between Riemannian foliation and totally geodesic foliation of a manifold (see Fig. \ref{fig1}).
In particular, a totally geodesic foliation has the so-called tangential parallelism and the corresponding Lie algebra (the duals of the transverse parallelism and its Lie algebra of the Riemannian foliation \cite{Molino}). In the case under consideration the Lie algebra is abelian, and it was proposed in \cite{Park:2014qoa} that the abelian symmetry be associated with the gauge symmetry that allows the gauge-fixing of the lapse and shift.
In other words, the lapse and shift gauge symmetry should somehow be related to the action of group fibration that generates the ``time" direction (i.e., the tangential parallelism; see below). The gauge-fixing then corresponds to taking the quotient of the bundle by the group, bringing us to the holographic reduction.\footnote{As stressed in \cite{Park:2014tia}, the reduced theory is not a genuine 3D theory but still a 4D theory whose dynamics can be described through the hypersurface.}

The potential significance of this abelian algebra for the physics context becomes much clearer once the whole mathematical setup is reconstructed in the framework of 
jet bundle theory (see, e.g., \cite{Saunders,Fatibene,Mangiarotti} for reviews of jet bundle theory). {For example, the jet bundle setup brings to light that the abelian symmetry is a gauge symmetry.} Here, we expand the work of \cite{Park:2014qoa}\footnote{{The work of \cite{Park:2014qoa} was the result of the efforts, while initially unaware of any of the works of \cite{York:1972sj,Moncrief:1989dx,Fischer:1996qg,Gay-Balmaz:2014ena,Gerhardt:2012ku}, to come up with an independent assurance of the claim in \cite{Park:2014tia}. The central motivation and focus of the present work is to further justify the quotient procedure proposed in \cite{Park:2014qoa}, and we believe that the jet bundle setup achieves that.}
To recapitulate, the main goal of the present work is to reinforce the mathematical 
picture of the reduction in \cite{Park:2014qoa} by adding the element of jet bundle theory.
As for the jet bundle theory, it is the {\em relevance} of the jet bundle that we focus on. (No new result will be added to the jet bundle theory itself.)
Once the reduction is established (physically \cite{Park:2014tia} or mathematically \cite{Park:2014qoa} (and present work)), the renormalizability follows (see, e.g., \cite{Anselmi:2003xb}). The explicit counter-term analyses have been presented in
\cite{Park:2014noa} (and more recently in \cite{Park:2015ota}). (For that, the standard perturbative quantum field theoretic techniques were employed.) 
} and elaborate on the mathematical picture by adding another ingredient, a jet bundle. 
The key result is that the proposal in \cite{Park:2014qoa} that the abelian symmetry be associated with the gauge symmetry now is on firmer ground, and we have	a refined confirmation of the mathematical picture of the reduction presented in \cite{Park:2014qoa}. In particular, the modding-out procedure, which is central for the mathematical reduction picture but only qualitatively outlined in \cite{Park:2014qoa}, is made quantitative and precise.

\vspace{.3in}

\ni The rest of the paper is organized as follows
\vspace{.1in}

\ni In the next section we start with an overview. We first compare the present approach with some of the existing approaches.
Because the present work employs the mathematical machinery of advanced differential geometry, we also present a pictorial (but logical) account before engaging with the formal mathematics. 
The review of necessary elements of differential geometry in section 3 covers the theory of foliation (that has been fruitful \cite{Park:2013vpa}) and jet bundle. The role of the jet bundle theory will be mostly conceptual in this work. The relevance of the jet bundle is that it is the object, as we will see below, whose group quotient yields the degrees of freedom of the first order Lagrangian,\footnote{
There exists a subtlety in conducting the quotient procedure; this subtlety lies in the fact that we are using the Palatini formalism, which in turn is due to our preference for employing the first (instead of the second )- order jet bundle.
As we will comment later, use of the second-order jet bundle theory would render the use of the Palatini formalism unnecessary as well as remove the subtlety.} and this is crucial for justifying the proposal in \cite{Park:2014qoa} of the abelian algebra as the gauge symmetry.
 Conceptual though it may be, the addition of the jet bundle theory will make the whole mathematical picture far more geometric and clearer than otherwise. The addition of the jet bundle theory will also have several interesting implications that we will discuss in the final section.
In section 4, which is the main part of the present work, we refine and expand the analysis of \cite{Park:2014tia}\cite{Park:2014qoa} with the combined setup of foliation and jet bundle theories. We first review the reduction by following a path slightly different from \cite{Park:2014tia} in section 4.1. 
The accounts in section 4.1 and \cite{Park:2014tia} are both within the conventional physics framework. The complementary and more mathematical account is given in section 4.2 where we explain precisely how the quotient by the abelian group associated with the totally geodesic foliation should be performed. {Taking the quotient is the step that corresponds to defining the physical states by a lapse function constraint in the physical analysis, a central step that led to reduction and renormalizability. The quotient was first suggested in \cite{Park:2014qoa} in a qualitative, less explicit manner. With the addition of the jet bundle, the meaning of the quotient becomes much clearer and quantitative.} We conclude with discussions of several issues and future directions in section 5. In particular, we present a preliminary discussion of quantization around a Schwarzschild background as one of the future directions.
Appendix A and B contain a concise review of the result in the jet bundle and derivation of \rf{constronn} needed in the main body.

\section{Overview}

Since the claim of renormalizability in \cite{Park:2014tia} is quite nontrivial we recapitulate some of the efforts made to substantiate the claim. To that end we first compare and contrast the approach of \cite{Park:2014tia} with the standard canonical quantization. As stated in the introduction the approach of \cite{Park:2014tia} has ingredients of the covariant quantization as well; we highlight what allows one to draw a conclusion different from the covariant method when it comes to renormalizability. We also comment on perspectives related to loop quantum gravity \cite{Thiemann:2007zz}.  

In \cite{Park:2014qoa}, the abelian Lie algebra associated 
with the totally geodesic foliation was envisaged as the gauge symmetry of the system, which was essential for the mathematical picture of the reduction. The main goal of the present more systematic and refined account of \cite{Park:2014qoa} is to set the view of the abelian Lie algebra as the gauge symmetry (and thereby the reduction picture itself) on firm mathematical ground. As we will see in section 4, the modding-out procedure outlined in \cite{Park:2014qoa} can be made precise with the setup of the jet bundle. However, the required element of jet bundle theory is abstract and perhaps not too familiar to physicists. For this reason it will be desirable to have a more informal way of understanding the mathematical picture: we present a pictorial account in the second subsection below.

\subsection{Comparison with existing approaches}

Various reductions in degrees of freedom were reported in the past in \cite{York:1972sj,Moncrief:1989dx,Fischer:1996qg,Gay-Balmaz:2014ena,Gerhardt:2012ku}, with which the works of \cite{Park:2014tia,Park:2014qoa,Park:2014noa} share certain features. 
All of these works employed the usual 3+1 splitting in which the genuine time coordinate was separated out. After expanding on the fact that the hypersurface can serve as a transverse and traceless spin-two representation of gravity, a conclusion was drawn in \cite{York:1972sj} that the spacelike hypersurface specified up to a conformal factor can be taken as the true degrees of freedom. 
In \cite{Moncrief:1989dx,Fischer:1996qg} Hamiltonian reduction was carried out on a class of 4D manifolds with certain topological restrictions and it was shown that the reduced Hamiltonian was given by the volume of the hypersurface. Intensive and extensive use was made of jet bundle theory in \cite{Gay-Balmaz:2014ena}. Reduction of a 4D Lagrangian to 3D was established with a certain crucial assumption in a general field theory context. These works concern the issue of the true degrees of freedom but did not directly address quantization of gravity. To address quantization, one should deal with the constraints, the so-called spatial diffeomorphism and Hamiltonian constraints.   
(We have called these constraints the shift vector and lapse function constraints, respectively, in the context of the Lagrangian analysis.) In our recent works \cite{Park:2014tia,Park:2014qoa}, we employed a 3+1 splitting with one of the spatial directions separated out. The strategy for reduction was clearly spelled out: removal of all of the unphysical degrees of freedom from the external states. We have addressed the quantization issue by solving the constraints. (A solution of the diffeomorphism constraint - which has been one of the major obstacles in gravity quantization - was previously attempted in \cite{Gerhardt:2012ku}.) The implication of the solution of the shift vector constraint has been brought out in \cite{Park:2014qoa} by foliation theory. An explicit (and slightly different) procedure of quantization has been presented in subsequent works \cite{Park:2014noa,Park:2015ota,Park:2015xoa}.

Let us recall the obstacle to renormalizability in the conventional covariant approach. To avoid unnecessary complications we mostly restrict to the pure gravity (i.e. no matter) in a flat background. Cancellation of various loop divergences requires counter-terms constructed out of the Ricci scalar, Ricci tensor and Riemann tensor. Suppose that somehow only the Ricci scalar and tensor but not the Riemann tensor were required. One can then introduce a metric field redefinition to absorb the counter-terms \cite{'tHooft:1973us} (this is possible because the Ricci scalar and tensor are proportional to the field equation; as well known, counter-terms proportional to the field equations can be absorbed by field redefinitions in general) and make the theory renormalizable. (The renormalizability is more complicated than usual cases in the sense that it requires a field redefinition unnecessary in a simply renormalizable theory in which only shifts in the parameters are required.)     
Although the Riemann-tensor-containing terms {\em do} appear among the counter-terms, the reduction to 3D makes it possible to reexpress it in terms of the Ricci scalar and tensor.
One may say, therefore, that the reduction is crucial to renormalizability. The reduction was originally proposed in \cite{Park:2014tia} and is reviewed in section 4.2 in its slightly modified form. The key observation for the reduction was the fact that the residual 3D gauge symmetry can be employed to gauge away the non-dynamical fields such as the lapse function and shift vector (or the matter fields when present). This is in additon to the standard bulk gauge-fixing. A detailed discussion of the 3D residual symmetry can be found in \cite{Park:2014noa}.
The actual procedure of renormalization was also initiated in \cite{Park:2014tia} where the importance of using the traceless propagator obtained by gauge-fixing of the trace piece was noted. The gauge-fixing of the tracepiece, which is one of the unphysical modes, was more systematically addressed in \cite{Park:2015ota} for a flat background and is generalized in \cite{Park:2015xoa} to a black hole background. A refined background field method was introduced in \cite{Park:2014noa} and further applied in \cite{Park:2015ota} to compute various one-loop counter-terms.\footnote{More recently, the method of \cite{Park:2014tia} has been applied to de Sitter background and a flat background with matter \cite{Park:2015ybl} after the reduction mechanism was substantially generalized.}

The method of \cite{Park:2014tia} has an undesirable feature: it has only 3D covariance (at the final stage once the physical state condition is enforced in, say, the 1PI effective action). However, what the approach is to the (future 4D) covariant approach should be what the lightcone quantization is to the BRST quantization, e.g., in string theory. In addition, the physical states have not been constructed in terms of gauge-invariant operators, unlike in loop quantum gravity. Dealing with gauge-invariant objects has certain advantages. (For example, gauge choice independence is guaranteed; pathology associated with incomplete gauge-fixing could be avoided, at least in principle) Again the approach of \cite{Park:2014tia} seems analogous to the lightcone quantization of string theory: while it should be a quickest way to obtain the physical spectrum it may be less elegant than more sophisticated methods. 

It should presumably be possible to translate the reduction observed in \cite{Park:2014tia} into the language of loop quantum gravity (which may add additional insights to, e.g., the task of unraveling the identity of the states constrained by diffeomorphism and Hamiltonian constraints within the formalism of loop quantum gravity). Once this is done the present approach may benefit (and thereby acquire higher level of elegancy) from the gauge-invariant ingredient of loop quantum gravity. 
Also, at some point in the development of the present approach, the formalism of the boundary state of loop quantum gravity \cite{Modesto:2005sj} may be useful. Eventually it would be desirable to quantize gravity in a gauge invariant and fully covariant manner.

\subsection{Reduction: pictorial account}

It was shown in \cite{Park:2014tia} that the physical states relevant for quantization around a flat background can be described by the 3 by 3 ``spatial" part of the original 4D metric with only 3D-coordinate dependence. (The reduced theory is not a genuine 3D theory in the sense that, unlike a genuine 3D theory, there are two surviving degrees of freedom inherited from the 4D theory.) The result was obtained through the field theoretic techniques. One can have a complementary view of the reduction by combining the foliation and jet bundle theories.

Jet bundle theory has been extensively used in the configuration reduction approach.
The concept of a jet is a generalization of a tangent vector. 
A vector tangent to a manifold can be defined as an equivalence class of curves passing through a point. A first order jet, which we will mostly consider in this work, is defined as an equivalence class of sections that have the same first order Taylor expansion at the point under consideration. 

We point out two facts about a jet bundle to motivate its use.\footnote{{ See also \cite{Esposito:1995vc,Rajpoot:2014fwa,Forger} and references therein for its use.}}
Consideration of tangent vectors is sufficient for the quantum mechanical variational calculus of particle dynamics. Once one considers a (quantum) field theoretic system, the jet bundle provides a naturally generalized setup for the variational calculus. The second fact for the relevance of jet bundle theory is gauge symmetry-related. For a reason that will become clear, we employ the Palatini
formalism of general relativity which is also called the first order formalism. The degrees of freedom of the Palatini formalism are the connection fields (and the metric). The connection in mathematics is usually defined as a map that projects the tangent space of the principal bundle onto the horizontal (or vertical) subspace of the tangent space. There is another way of realizing the connection and that is in terms of the jet bundle: a connection can be viewed as a section of the first order jet bundle.
Consider the jet bundle $J^1P$ (a formal introduction will be given in one of the appendices; for now it is sufficient to view it as a certain bundle over the base $P$, the total space of a principal bundle).
The connections that we need belong to a special class of connections called the principal connections. The principal bundle relevant for us is the linear frame bundle of the 4D manifold $M_{4D}$, i.e., $P=LM_{4D}$. As we will review below, one gets a principal connection (and the connection fields, the degrees of freedom of the Palatini formalism) 
once $J^1P$ is modded out by the gauge group $G$, which is $GL(4)$ in the present case. More generally, the configuration space for an arbitrary principal bundle is given by modding the jet bundle by the gauge group, $J^1P/G$. It is well known in the mathematical literature that the resulting space is the space of connections, ${\cal C}P$, namely,
\bea
{\cal C}P=J^1P/G
\eea
In other words, the bundle $J^1P$ is the object whose group quotient yields the degrees of freedom of the first order Lagrangian. (Of course, a gauge-fixing - which corresponds to choosing a local trivialization - is to be performed on these degrees of freedom.)

\begin{figure}
\centerline{
\begin{minipage}[b]{12 cm}
             \epsfxsize=14cm
              \epsfbox{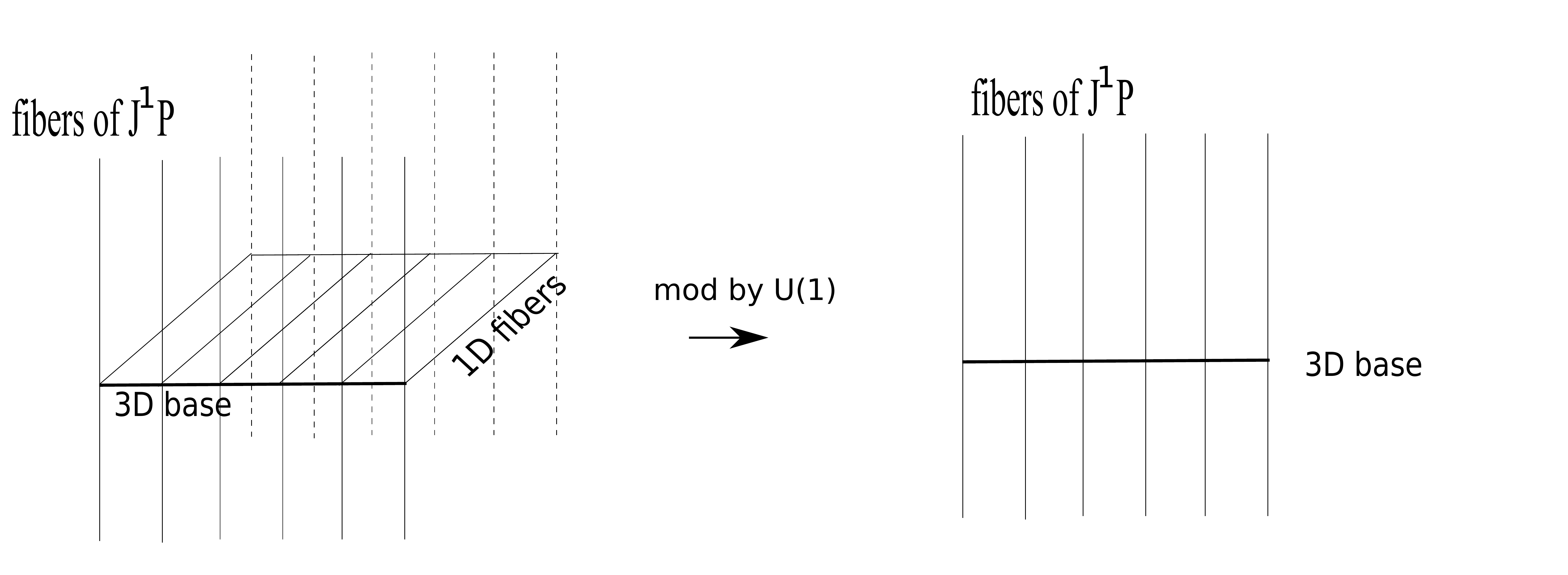}
      \end{minipage}
      }
\caption{bundles }
\label{fig}
\end{figure}

The Riemannian foliation condition obtained in \cite{Park:2014tia} implies that the original 4D manifold itself - which is the base for the frame bundle - can be viewed as an abelian {\em principal bundle} over a 3D base $B_{3D}$ with geodesic fibers. In other words, the original 4D manifold $M_{4D}$ can be seen as the total space of a $U(1)$ principal bundle. The central issue then is what to do with this abelian gauge symmetry.

The extra abelian gauge symmetry has arisen from constraining the configuration space. Motivated by the physical analysis in \cite{Park:2014tia} and the fact that the $U(1)$ symmetry should be a gauge symmetry, it was proposed in \cite{Park:2014qoa} that the configuration space be modded out by the $U(1)$ symmetry. Given that the unconstrained configuration space corresponds to $J^1LM_{4D}/GL(4)$,  we propose in section 4 that the quotient by $U(1)$ should be such that it reduces the 4D bundle $M_{4D}$ to its 3D base $B_{3D}$.\footnote{In essence, it must be possible to understand the modding-out procedure along the line of the symplectic quotient of the configuration reduction approach \cite{Cendra,Marsden,Gay-Balmaz:2014ena} once the relevane of the totally geodesic foliation is understood.} Then the whole jet bundle will reduce to $J^1LB_{3D}/GL(3)$, as shown in the figure.

\section{Foliation and jet bundle}

In this section, we review necessary elements of differential geometry \cite{Kobayashi1,Kolar,Walschap,Petersen,Goldberg} and significantly strengthen  the mathematical foundations laid out in \cite{Park:2014qoa} in support of the observation in \cite{Park:2014tia}. 
The needed elements of differential geometry are foliation and jet bundle theories. 
Below, we will make some efforts to relate various mathematical content to the physics contexts.

One of the things that we will need in the analysis below is a commutator of the Lie derivative and covariant derivative.
By using the definitions of the Lie and covariant derivatives,
one can show the following relation \cite{Kobayashi1}:
\bea
[\mathscr{L}_{\bfX}, \N_{\bfY} ]{\bf T}=\N_{[\bfX,\bfY]}{\bf T} \la{liecocom}
\eea
where $\bfX, \bfY$ are vector fields and ${\bf T}$ is a tensor field. The bold-faced letters represent the coordinate-free quantities. As commented in \cite{Kobayashi1} it is only necessary to prove \rf{liecocom} when ${\bf T}$ is a scalar or vector: the general tensor case follows.
We illustrate the proof by taking ${\bf T}$ to be another vector field ${\bf T}={\bf Z}$:
\bea
\mathscr{L}_{\bfX}\N_{\bfY} {\bf Z}&=&
\lim_{t\ra 0}\fr{\N_{\vf_t(\bfY)}[ \vf_t({\bf Z})]-\N_{\bfY} {\bf Z}}{t}  \nn\\
&=&\lim_{t\ra 0}\fr{\N_{\vf_t(\bfY)}[ \vf_t({\bf Z})]
                -\N_{\vf_t(\bfY)} {\bf Z}
 +\N_{\vf_t(\bfY)} {\bf Z}-\N_{\bfY} {\bf Z}}{t}  \nn\\
&=&\N_\bfY[{\mathscr{L}_{\bfX}}{\bf Z}]+\N_{\mathscr{L}_{\bfX}\bfY}{\bf Z}\nn\\
&=& \N_\bfY {\mathscr{L}_{\bfX}}{\bf Z}+\N_{[\bfX,\bfY]}{\bf Z} \la{liecocompr}
\eea
where we have used the identity $\mathscr{L}_{\bfX}\bfY=[\bfX,\bfY]$ in the last equality; $\vf_t$ denotes a one-parameter group of diffeomorphisms.

\subsection{Riemannian vs. totally geodesic foliation}

We highlight in this section that there are two special types of foliation that are dual. (General aspects of foliation theory can be found in, e.g., \cite{Molino,Moerdijk,Gromoll,Rovenskii,Candel}.) They are the so-called Riemannian foliation (also called metric foliation) and totally geodesic foliation.  
We will see that the condition obtained in \cite{Park:2014tia} by examining the shift vector constraint can be interpreted as the condition for the foliation of $M_{4D}$ to be Riemannian. A Riemannian foliation admits the totally geodesic foliation and vice versa. The connection between gauge-fixing and reduction becomes clearer in the context of the totally geodesic foliation.

A globally hyperbolic spacetime that we focus on in this work\footnote{{ As will be discussed later, focus on a globally hyperbolic spacetime is a matter of convenience but not a strict requirement.}} admits a foliation of a family of hypersurfaces $\Sigma_{x^3}$ with the base manifold parameterized by a ``time" coordinate $x^3$.  
Let us choose a coordinate system such that a vector $\bfX$ takes
\bea
\bfX\equiv {\bf\pa}_\m =(\pa_{x^3},\pa_m),\quad m=0,1,2
\eea
The vector $\pa_{x^3}$ can be decomposed according to
\bea
\pa_{x^3}=n\hat{\bfn}+N^m \pa_m
\eea
where $\hat{\bfn}$ is the unit vector normal to the hypersurface and $\pa_m$ is a vector tangent to $\Sigma_{x^3}$. The fields $n$ and $N^m$ are called the lapse function and the shift vector respectively.
The components of the metric tensor $g_{\m\n}\equiv \bfg(\pa_\m, \pa_\n)$ are given in the conventional notation by
\bea
ds^2=g_{\a\b} dx^\a dx^\b=(n^2 +\g^{mn}N_mN_n)(dx^3)^2+N_m dx^3dx^m+\g_{mn}dx^m dx^n
\eea
where the far right-hand side is the ADM parameterization of the metric components.
The foliation has 3D leaves with metric denoted by $\g_{mn}$. The space of the leaves (i.e., the base manifold) is parameterized by $x^3$.

\begin{figure}
\centerline{
\begin{minipage}[b]{12 cm}
             \epsfxsize=12cm
              \epsfbox{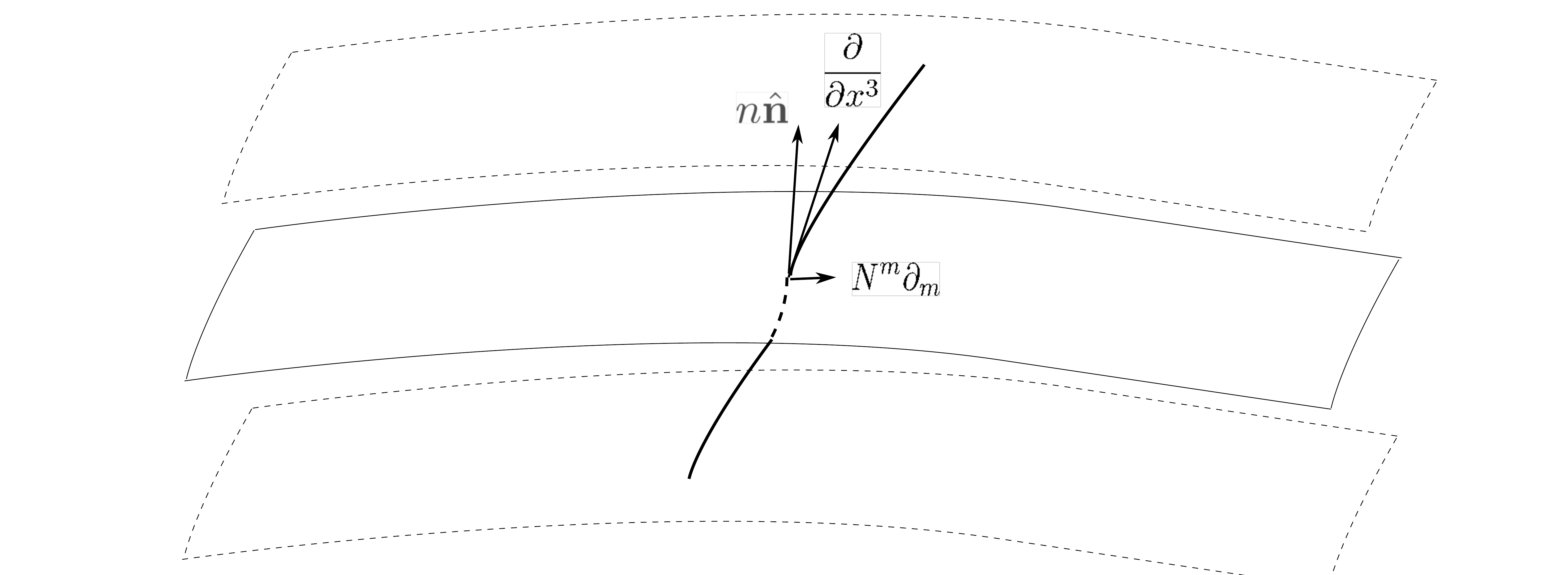}
      \end{minipage}
      }
\caption{lapse function and shift vector}
\label{fig2}
\end{figure}

Intuitively, a Riemannian foliation is such that the metric on a leaf does not change as one moves along the transverse direction(s), a property called the transverse parallelism. A precise definition of Riemannian foliation can be introduced in terms of the 
horizontal component of the metric tensor, $\bfg^h$: given $\bfX,\bfY \in \mathfrak{X}(M)$ ($\mathfrak{X}(M)$ denotes the set of all vector fields on a manifold $M$),
\bea
\bfg^h(\bfX,\bfY)\equiv \bfg(\bfX^h,\bfY^h)
\eea 
where $\bfX^h,\bfY^h$ denote the horizontal component of $\bfX,\bfY$. The necessary and sufficient condition for the foliation to be Riemannian is
\bea
\mathscr{L}_{\bf Z} \,\bfg^h=0,\quad {\bf Z}\in {\cal V} \la{rfcon}
\eea
This condition is a statement about invariance of the horizontal metric under the flow of vertical vector
fields.

It is known in the mathematical literature that the Riemannian foliation admits the dual totally geodesic foliation. A totally geodesic foliation is characterized in terms of the second fundamental form.
The second fundamental form $\bfK$, also called the extrinsic curvature, 
of a given hypersurface $\Sigma_{x^3}$ is defined by
\bea
\bfK(\bfA,\bfB)=\bfg(\bfA,\N_\bfB\, \hat{\bfn})
\eea
where $\bfA, \bfB$ represent the vectors tangent to the hypersurface.
In a local coordinate system, this takes
\bea
K_{mn}=(\N_\n n_\m) e_m^\m e_n^\n=\fr12 (\mathscr{L}_{n}g_{\m\n})  e_m^\m e_n^\n
\eea
where 
\bea
e_m^\r\equiv \fr{\pa x^\r}{\pa y^m}
\eea
in terms of the bulk coordinates $x^\m$ and hypersurface coordinate $y^m$.
After a series of manipulations, one can show that $K_{mn}$ can be expressed in the form given in \rf{K4defqq} below in terms of the lapse function and shift vector.

The aspect of foliation theory most relevant for the present work is the parallelism(s) of the totally geodesic (and Riemannian) foliation(s). Basically, the parallelism of a totally geodesic foliation is a set of vector fields of ${\cal F}$, the leaves. The number of the vector fields is equal to the dimensions of a leaf; it is one for the foliation under present consideration.

\subsection{Connection as section of jet bundle}

The usual definition of a connection as a projection map on the tangent bundle $TP$ is useful for certain purposes. However, the realization of a connection as a section of a jet bundle is more convenient for certain other purposes.  
 The first order jet, which we will focus on, is a generalization of a tangent vector (recall that a vector tangent to a manifold can be defined as an equivalence class of curves passing through a point): it is defined as an equivalence class of sections that have the same first order Taylor expansion at the point under consideration. The first jet bundle of a principal bundle yields, upon being modded out by the gauge group, the bundle of the principal connections. Here we present a short introduction to a jet bundle and how a connection is realized in the jet bundle context.

A second order jet bundle might be more suitable for describing the usual form of the Einstein-Hilbert action. Instead of turning to the second jet bundle, we will consider, for the mathematical analysis in section 4.2, the Palatini formulation of general relativity in conjunction with the first order jet bundle. The Palatini formulation is also called the first order formulation, and  
consideration of the {\em first order} jet bundle is sufficient. 
The degrees of freedom of the Palatini action are the metric components and the connection fields that become identified with the Christoffel symbol after use of their field equations.

The result in jet bundle theory that will be used to build the geometric picture in section 4 is the well-known fact:
\bea
J^1P/G={\cal C}P  \la{conn}
\eea
where $J^1P/G$ represents the quotient by the group $G$ of the first order jet bundle of the principal bundle under consideration, $J^1P$. The right-hand side, ${\cal C}P$, stands for the bundle of the principal connections.  
In Appendix A the relation above will be derived by mostly following \cite{Fatibene}; the relation \rf{conn} will be applied to the reduction scheme that we propose in the next section.

\section{ Reduction: analytic account}

In this section, we put together all the ingredients reviewed in the previous sections, and present the way in which the reduction takes place in the resulting setup.\footnote{{ In this work we do not keep the mathematical rigor at the same level required for mathematics literature. For example, the transverse parallelism of Riemannian foliation was originally announced for a compact manifold \cite{Molino}. Although the same is true for a noncompact manifold, some of the nice properties of Riemannian foliation are only for a compact manifold. (See, e.g., \cite{Nozawa}.) Generalization of the Molino's result to noncompact cases seems to be a fairly recent research topic in foliation theory. Also, we are cavalier about the difference between a Riemannian manifold and a Lorentzian manifold.}} More precisely speaking, we reinforce and make precise the mathematical picture of \cite{Park:2014qoa} by the jet bundle theory. This is done by providing a mathematical foundation to the view in \cite{Park:2014qoa} of the abelian parallelism as the gauge symmetry.

In section 4.1, we present an analysis of the ADM Lagrangian that is slightly different from that of \cite{Park:2014tia} in that here the $\g^{mn}$ field equation is utilized. The ADM formalism employs the 3+1 splitting of the coordinates, and it is usually the time coordinate that is separated from the rest. The focus of \cite{Park:2014tia} and the present work is separation of one of the {\em spatial} coordinates \cite{Sato:2002kv}. 
If one considers the genuine time direction (as opposed to, say, a radial coordinate) to be a split direction, then it is physical to consider a globally hyperbolic spacetime since a globally hyperbolic spacetime has causally nice properties. (Since most of the interesting spacetimes are globally hyperbolic and the restriction to a globally hyperbolic spacetime is of a topological nature from a mathematical point of view, it will be a very mild restriction.) However, the meaning of a ``globally hyperbolic spacetime" is only an analogy if the split direction is spatial although the techniques that we will employ apply regardless. As a matter of fact, the condition of a ``globally hyperbolic spacetime" is not strictly necessary.\footnote{{ However, there is one place in the analysis of \cite{Park:2014noa} where restriction to a globally hyperbolic spacetime might be required.}} All we need is the condition that the spacetime is of Riemannian foliation; then the dual totally geodesic foliation will follow.

Although quantization of gravity in a flat background was considered in \cite{Park:2014tia}, it should be possible to apply the method to a class of relatively simple backgrounds such as stationary black holes. The method should also be applicable to a de Sitter spacetime in the static coordinates. (It might be applicable to an anti-de Sitter spacetime as well, as a preliminary analysis indicates. However, the usual issue of whether or not a S-matrix exists is expected to cause complications.) The method may not be applicable (at least not in a straightforward manner), e.g., to a generic time-dependent background for the reason that we will comment on below. The ADM formulation may not be convenient and/or suitable for such a manifold anyway (see, e.g., \cite{Frolov:2008sn} for a related discussion.)

In section 4.2, we will explore where the combined inputs of physics and mathematics take us once the totally geodesic foliation is utilized in the jet bundle setup. 
As first commented in \cite{Park:2014tia}, the extra abelian gauge symmetry arises from constraining the configuration space through the shift vector constraint. Motivated by the physical analysis in \cite{Park:2014tia} and the fact that the $U(1)$ symmetry should be a gauge symmetry, it was proposed in \cite{Park:2014qoa} that the configuration space be ``modded out" by the $U(1)$ symmetry as well as the usual quotient by the symmetry group. 
Compared with \cite{Park:2014qoa}, a new ingredient is the jet bundle.
 We propose the precise way in which the quotient by $U(1)$ is to be performed and explain the new insights offered by the jet bundle.

\subsection{Analysis in the ADM setup}
In this section, we review the quantization procedure in a flat background \cite{Park:2014tia}.
Consider the 4D Einstein-Hilbert action 
\bea
S=\int d^4 x \sqrt{-g}\;R  \la{unsplit}
\eea
and quantization in the operator formalism. 
We split the coordinates into
\bea
x^\m\equiv (y^m,x^3) \la{coord}
\eea
where $\m=0,..,3$ and $m=0,1,2$. (One can choose the Cartesian coordinate system $x^\m=(t,x,y,z)$ with $x^3\equiv z$ for a flat background.)
By parameterizing the 4D metric \cite{Arnowitt:1962hi}\cite{Poisson} in the the 3+1 split form, one gets 
\bea
g_{\m\n}=\left(
\begin{array}{cc}
\g_{mn} & N_{ m} \\
&\\
N_{ n} &  n^2+\g^{mn}N_{m} N_{ n} 
\end{array}
\right)\quad,\quad
g^{\m\n}=\left(
\begin{array}{cc}
\g^{mn}+\fr1{n^2}N^m N^n & -\fr1{n^2}N^{ m} \\
&\\
-\fr1{n^2}N^{ n} &  \fr1{n^2} 
\end{array}
\right)
\eea
where $n$ and $N_{m}$ denote the lapse function and shift vector respectively.
For a flat background, which we are considering, the background part of the lapse function, $n_0$, takes $n_0=1$. As we will show the fluctuation part is absent in the gauge chosen below.
The action takes
\bea
S=\int d^4 x\;n\sqrt{-\g} \left(R^{(3)}+K^2-K_{mn}K^{mn}\right)
\la{1p3act}
\eea
with the second fundamental form given by
\be
K_{mn}=\fr1{2n}\left(\mathscr{L}_{\pa_{3}} \g_{mn}-{\nabla}_m N_{n}
         -{\nabla}_n N_{ m} \right),\qquad K=\g^{mn}K_{mn}.
\la{K4defqq}
\ee
where $\mathscr{L}_{\pa_{3}}$ denotes the Lie derivative along the vector field $\pa_{x^3}$ and $\N_m$ is the 3D covariant derivative constructed out of $\g_{mn}$. The shift vector and lapse function are non-dynamical and their field equations are 
\bea
{\N}_m (K^{mn}-\g^{mn} K)=0  \la{Ncon}
\eea
\bea
R^{(3)}-K^2+K_{mn}K^{mn}=0  \la{ncon}
\eea
{These should be imposed as constraints that are the analogues of the spatial diffeomorphism and Hamiltonian constraints of the Hamiltonian analysis. Although we have a perturbative analysis in mind in the later stage\footnote{The perturbative analyses have recently been carried out in \cite{Park:2014tia,Park:2014noa,Park:2015ota,Park:2015xoa}.}, most of the equations below will be kept at the full non-linear level. (Physical significance of some of the equation will be clearer in the non-linear form.) Nevertheless, one should envisage in several places the splitting of a field $\Phi \;(\equiv n,N_n,\g_{mn})$ into the background and fluctuation:
\bea
\Phi\equiv \Phi_0+\tilde{\Phi}
\eea
}
The 4D bulk gauge symmetry can be fixed by imposing the de Donder gauge. The bulk gauge fixing leaves 3D residual gauge symmetry (see the discussion in \cite{Park:2014noa}); by using this 3D symmetry, the shift vector can be gauged away
\bea
N_{m}=0,   \la{Nazero}
\eea
As far as the perturbative scattering analysis, say, in \cite{Park:2014noa} is concerned, this gauge-fixing means that the fluctuation part of $N_m$ is removed. (Note that the flat spacetime is such that the background part of $N_m$ is absent.)
Substitution of $N_{m}=0$ into \rf{Ncon} leads to
\bea
{\N}^m \left[\fr{1}{n}\Big(\mathscr{L}_{\pa_3} \g_{mn}
 -\g_{mn}\g^{pq}\mathscr{L}_{\pa_3} \g_{pq}\Big)\right]=0  \la{mtmconstr}
\eea
which in turn implies\footnote{For a generic time-dependent background, there are several things that may prevent the present method from being applicable. First of all, the ADM formalism may not be convenient or suitable for such a spacetime. Even if it is, there is a question of whether or not the fixing in \rf{Nazero} would be legitimate. Not unrelated to this, a coordinate system consistent with \rf{constronn} must exist for the applicability.} \cite{Park:2014qoa}
 \bea
 \pa_m n=0 \la{constronn}
 \eea
Its derivation has been quoted in the appendix for convenience. (This with the non-dynamism of $n$ implies that one can set $n=n_0=1$ {(this corresponds to the gauge-fixing of the fluctuation part, $\tilde{n}$, of $n\equiv n_0+\tilde{n}$ with the background part $n_0$ set to its flat background value, $n_0=1$.)}
 The condition \rf{constronn} can be rewritten as
\bea
\mathscr{L}_{\pa_m} n=0  \la{Riemannconp}
\eea
This is precisely the condition \rf{rfcon} for the foliation of $M_{4D}$ to be Riemannian.
(Note that the lapse function $n$ corresponds to the $(x^3,x^3)$ component of $\bfg^h$.)
 
We will come back to this below but for now let us consider the form of the action \rf{1p3act} and re-derive, by using a slightly different method, one of the equations obtained in \cite{Park:2014tia}:  
\bea
K_{pq}K^{pq}=0 \la{Kpqsq}
\eea  
which implies reduction of the 3 by 3 sector of the 4D metric components to 3D. In the next subsection, we will compare the way in which this reduction has taken place with the corresponding reduction in the mathematical analysis based on the jet bundle setup.

Let us first recall that the full nonlinear bulk de Donder gauge $g^{\r\s}\G^\m_{\r\s}=0$ \cite{Smarr:1978dia} reads, in terms of the ADM fields,
\[
 \!\!\!\!(\pa_{x^3}-N^m \pa_m) n=n^2K  
\]
\bea
\;\;(\pa_{x^3}-N^n \pa_n)N^m=n^2(\g^{mn}\pa_n \ln n-\g^{pq}\G^m_{pq})
\la{ADMdd}
\eea
On account of the first equation, the constraint \rf{constronn} implies
\bea
K=0
\eea
The $\g_{mn}$ field equation of \rf{1p3act} can be obtained by taking the variation with respect to $\g^{mn}$ (we have set $n=1, N_m=0$)\footnote{{ For a black hole background, this condition and the nonlinear de Donder gauge conditon \rf{ADMdd} should be modified. We will expound on this in the discussion section.}}:
\bea
&&R_{mn}^\3-\fr12 \g_{mn}R^\3\nn\\
&&-\fr12  \g_{mn}K^2
+\fr1{\sqrt{\g}}\g_{pm}\g_{qn}\mathscr{L}_{\pa_{3}}(\sqrt{\g}\;K\g^{pq})
+K\mathscr{L}_{\pa_{3}}\g_{mn}  \nn\\
&&+\fr12  \g_{mn}K_{rs}K^{rs}-2 K_{mp}K^p{}_n
-\fr1{\sqrt{\g}}\g_{mp}\g_{nq}\mathscr{L}_{\pa_{3}}(\sqrt{\g}\;K^{pq})\nn\\
&&=0   \la{gmneom}
\eea
where the second line and third line come from $\d(\sqrt{\g}K^2)$ and $\d(\sqrt{\g} K_{pq}K^{pq})$, respectively. By using
\bea
\fr1{\sqrt{\g}}\g_{pm}\g_{qn}\mathscr{L}_{\pa_{3}}(\sqrt{\g}\;K\g^{pq})
=\g_{mn}\pa_3 K-2 K_{mn}K+\g_{mn}K^2
\eea
and a similar expression for $\fr1{\sqrt{\g}}\g_{mp}\g_{nq}\mathscr{L}_{\pa_{3}}(\sqrt{\g}\;K^{pq})$,
the field equation \rf{gmneom} simplifies to
\bea
&& R^\3_{mn}-\fr12\g_{mn}R^\3+\fr12  \g_{mn}K^2 +\g_{mn}{\pa_{3}}K
  +\fr12  \g_{mn}K_{rs}K^{rs} \nn\\
&&-2 K_{mp}K^p{}_n - K_{mn}K
-\g_{mp}\g_{nq}{\pa_{3}}K^{pq}=0
\label{gmneomflat}
\eea
Multiplication of $\g^{mn}$ yields
\bea
-\fr12 n R^\3+\fr12K^2-\fr12  K_{rs}K^{rs}+3\pa_3 K-\g_{pq}\pa_3 K^{pq}=0
\eea  
Combining this with the $n$-field constraint \rf{ncon} implies
\bea
&&3\pa_3 K-\g_{pq}\pa_3 K^{pq}
=2\pa_3 K+2 K_{pq}K^{pq}=0
\eea 
One of the de Donder gauge conditions yields $K=0$, hence we arrive at \rf{Kpqsq} through a different route:
\bea
K_{pq}K^{pq}=0,
\eea
{To paraphrase, we have combined the lapse function constraint with a field equation and a gauge-fixing condition; this step is a part of the legitimate procedure of determining physical states.} 
This way of reaching the reduction is complementary to that of \cite{Park:2014tia} and reassures the analysis therein. An issue that concerns the full nonlinear form of the de Donder gauge \rf{ADMdd} is worth noting.  
The form \rf{ADMdd} was imposed for quantization around a flat background, which was the focus of \cite{Park:2014tia}. The gauge condition should be modified once one considers quantization around a Schwarzschild black hole background.  
This can be seen from the fact that a Schwarzschild black hole background does not satisfy \rf{ADMdd} (whereas a flat background does). One should investigate whether it is possible, say, to modify the de Donder gauge appropriately in order to establish the reduction in the case of a Schwarzschild black hole.
We will have more on this in Discussion.

\subsection{Reduction via totally geodesic foliation \la{main}}

The analysis of the previous subsection, most of which was first presented in \cite{Park:2014tia}, was carried out in the framework of the conventional techniques of (quantum) field theory, and did not employ the mathematical duality between the Riemannian and totally geodesic foliations. A more mathematically oriented viewpoint was taken in \cite{Park:2014qoa}. The addition of the jet bundle element not only furnish the view of the abelian algebra as the gauge symmetry with a methematical foundation but also gives enlightening geometric insights to the whole picture as we will now discuss.

Let us start with a 4D manifold $M_{4D}$, and consider a $GL(4)$ principal bundle $LM_{4D}$, the bundle of linear frames.\footnote{{Strictly speaking, the connection that appears in general relativity is an affine connection and therefore, one should consider the affine extension of the frame bundle. However, we will just consider a linear frame bundle because there is, as well-known, an isomorphism between the linear connections and affine connections.}} Consider the first jet bundle of this bundle; taking a quotient by $G$ will lead to the principal connection as reviewed in section 3. The connection fields will be four-dimensional; i.e., they will depend on the 4D coordinates at this point. In the analysis of the previous subsection we did not make use of the $U(1)$ gauge symmetry, but rather relied on various constraints to reach the reduction. It should be possible to repeat the analysis of \cite{Park:2014tia} that was reviewed above in the setup of a jet bundle in a more mathematically sophisticated manner, regardless of whether or not an advantage is taken of the existing totally geodesic foliation.

Regarding the approach in which the duality is not taken advantage of, one would start with the jet bundle $J^1LM_{4D}$ and consider
\bea 
{J^1LM_{4D}}/{GL(4)}
\eea
One would then proceed and gauge-fix some of the metric components. Although the abelian transverse parallelism is present after imposing the shift vector constraint, its presence is not a conspicuous feature of the system in this approach simply because one does not make use of it.
The final outcome will be the same: the physical configuration space will be reduced to the 3D base manifold.

It is through the approach in which the dual totally geodesic foliation is taken advantage of that the virtue of the jet bundle can really be appreciated. 
Let us now see how the dual totally geodesic foliation figures into the picture and leads to reduction of the connection fields to 3D.

In the dual picture, the original Riemannian foliation of codimension-1 can be viewed as the totally geodesic foliation of codimension-3. In other words, the manifold can be viewed as 1D fibration over the 3D base, and {the 1D fibration will be totally geodesic.} 
A totally geodesic foliation carries Lie algebra \cite{Cairns}, which is the so-called tangential Lie algebra (or the dual of the transverse Lie algebra); in the present case, the Lie algebra is abelian.

Before we spell out the precise quotient procedure, let us note that there exists a subtlety in conducting the procedure. The subtlety lies in the fact that we are using the Palatini formalism, which in turn is due to our preference for employing the first (instead of the second ) order jet bundle. The point is that the Riemannian foliation condition was obtained from the usual (i.e., second order) Lagrangian in the ADM form. 
Therefore the step of modding out by $U(1)$ seems justified only after making the Lagrangian partially on-shell by substituting the field equation of the connection field. But with this, the Lagrangian becomes second-order and seems potentially in conflict with the use of the first-order jet bundle. This seems to be a delicate issue\footnote{However, we do not believe that this issue is a genuine problem: we are just being absolutely careful. Moreover, use of the second jet will cure the problem as we will comment in the discussion.} and we will have more comments on this subtlety in the discussion section; for now, we set it aside and focus on the precise manner in which the procedure of the quotient should be performed on the first-order jet bundle.

The symmetry associated with $U(1)$ tangential parallelism is a {\em gauge symmetry} of the constrained space, therefore, the constrained configuration should be modded out by the symmetry. 
In other words, once the space is constrained by the shift vector constraint, one should consider (say, in the path integral) only the 4D manifolds $M_{4D}$s that are bundles of a $U(1)$ fibration over the 3D base.  
The flow of the $U(1)$ Lie algebra should be associated with the ``time" and viewing the $U(1)$ symmetry as a gauge symmetry is consistent, from this standpoint as well, with the well-known fact in general relativity that time-evolution is a gauge artifact.

Finally {\em the} central task: how the quotient procedure of the jet bundle is to be performed. Once one considers the constrained space, the modding-out by $U(1)$ should be carried out on the 4D manifold $M_{4D}$ itself. This will reduce $M_{4D}$ to its base $B_{3D}$. (We have used the fact that the quotient of $M_{4D}$ by $U(1)$ will be such that it reduces the 4D bundle $M_{4D}$ to its 3D base $B_{3D}$ since $M_{4D}$ itself can be viewed as a $U(1)$ principal bundle in the dual totally geodesic picture.)
Now with the original jet bundle reduced to the jet bundle over the 3D base, one gets the principal connection that is defined over the 3D base by taking the quotient by $GL(3)$ (see Fig. 1).

\section{Discussion}

In this work, we have expanded and refined the work of \cite{Park:2014qoa} by introducing jet bundle theory. The connection - defined as a map from the tangent space of the principal bundle to the horizontal (or vertical) space - can also be realized as a section of the first order jet bundle. The first order jet bundle has been considered in conjunction with the Palatini formalism of general relativity. 
The $U(1)$ symmetry should be modded out because it is now evident that it is a gauge symmetry of the reduced configuration (i.e., reduced by the shift vector constraint).
We have presented an enlarged geometric picture of the reduction: 
the reduced physical configuration space is obtained by considering $M_{4D}/U(1)=B_{3D}$ as the base space and considering the principle bundle, $LB_{3D}$ and the jet bundle, $J^1LB_{3D}$.

With the present mathematical version of the reduction, one may say that the reduction phenomena is {\em not} limited to a configuration governed by the Einstein's equation but is a more general phenomenon occurring {\em to a Riemannian manifold with a special foliation, the Riemannian foliation.}

In the main body, we have often referred to the analysis in \cite{Park:2014tia} as a physical approach and the one in \cite{Park:2014qoa} as a more mathematically oriented approach. In fact, it may not be a matter of a physical versus mathematical approach; a more proper view should be that they are the dual approaches in how one handles the quotient by $U(1)$.
It should be possible to repeat the analysis in \cite{Park:2014tia} in a jet bundle setup. 
One would start with $J^1LM_{4D}$ and consider
\bea 
{J^1LM_{4D}}/{GL(4)}
\eea
Gauge-fixing the lapse function and shift vector amounts, in effect, to carrying out the quotient by $U(1)$.

In the Palatini formalism, use of the the connection field equation yields the usual second order action. Therefore, modding out by $U(1)$ should be viewed as going on-shell (on-shell in the sense that the connection field equation is used). There should be a sense (on which our analysis in the main body has relied) in which using Palatini formalism and modding out by $U(1)$ can be taken as to ``commute". This is also indicated by the following fact. Instead of using the field equation to relate the Christoffel to the metric, rely on the metric compatibility condition which one imposes anyway.

\vspace{.1in}

There are several future directions\footnote{Some of these future directions have now been completed. For example, the quantization in a Schwarzschild background have recently been analyzed in \cite{Park:2015xoa}.}:

{It should be possible to achieve the reduction within the usual second order formalism that has been avoided for minimal formalism in this work. The setup based on the second jet bundle will not have the subtlety present in the setup of the first order jet. In that setup, one would consider $J^2LM_{4D}/G$. Presumably, this space will not form a new space but rather correspond, in the physical terms, to an extra derivative acting on the Christoffel symbols.\footnote{I thank T. Ratiu and L. Fatibene for sharing their expertise on this matter.} Here again the quotient procedure will reduce $M_{4D}$ to $B_{3D}$. Making this procedure more precise with an introduction of the second jet would be of some interest.}

 It should be of some interest to identify the $U(1)$ symmetry within the conventional Lagrangian field theoretic setup. This would require examination of various commutators (or Poisson/Dirac brackets) among the fields and constraints. Once identified, it will be possible to make the connection between the approaches of \cite{Park:2014tia} and \cite{Park:2014qoa} more concrete. {From various indications and the analysis so far (see \cite{Park:2015ybl} for a recent discussion), the abelian symmetry should be identified with the translation along the $x^3$ direction, the ``vertical" lines in Fig. \ref{fig1}(c). The quotient procedure corresponds to imposing a physical constraint such that the physical states are annihilated by the generator, the ``Hamiltonian." The crucial point that brings the reduction is that once the shift vector constraint is enforced, the Hamiltonian becomes identical to the lapse function constraint. The mathematical picture hinges on the condition of the original manifold being Riemannian. In this sense it indicates towards the possibility that the reduction may take place in cases in which a gauge different from the synchronous-type gauge is taken, and it would be interesting to explore the most general condition for reduction to occur.}

Extending the quantization analysis of \cite{Park:2014tia} and \cite{Park:2014noa} to a Schwarzschild background 
\bea
ds_{4D}^2=-\Big(1-\fr{2GM}{r}\Big)dt^2+\Big(1-\fr{2GM}{r}\Big)^{-1}dr^2+r^2(d\th^2+\sin^2\th d\varphi^2)
\eea
will be the primary interesting direction.
Although our focus has been quantization around a flat background, we believe the present method will apply to other relatively simple backgrounds such as a Schwarzschild's. The criterion for applicability of the present method will be whether or not the background under consideration is consistent with the shift vector constraint. Once the background is consistent with the constraint, we believe that the presence of the totally geodesic foliation suggests that reduction would take place one way or another. (The same seems to be strongly indicated by the configuration reduction approach \cite{Cendra}.)

 In the case of a Schwarzschild background, the radial direction will be separated out: $x^3\equiv r$. The quantization around this background is likely to be more subtle than the quantization around a flat background. One of the necessary steps should be a modification of the nonlinear form of the de Donder gauge $g^{\m\n}\G_{\m\n}^\r=0$ since the Schwarzschild background does not satisfy this condition. One may try the following modification,
\bea
g^{\m\n}\G_{\m\n}^\r=g^{\m\n}\G_{\m\n}^\r|_{g=g_0}
\eea 
where the right-hand side represents $g^{\m\n}\G_{\m\n}^\r$ evaluated at the Schwarzschild solution. At the linear level, this choice leads to the usual form of the curved space de Donder gauge used in perturbative analyses.   
Also, we do not expect to get $K_{mn}K^{mn}=0$ but instead an appropriately modified expression.

Finally, the quantization proposal of \cite{Park:2014tia} is subject, in principle, to the issue that any reduced space quantization faces: in general, gauge-fixing can be subtle. For example, it was noted that the reduced space quantization does not lead to the same physics as Dirac quantization \cite{Ashtekar:1982wv}\cite{Schleich:1990gd} (and refs therein). This was demonstrated by taking an example in which the first class constraint was not associated with the gauge symmetry. We are not aware of any work in which a similar conclusion was drawn for a theory with a gauge-symmetry inducing a first class constraint; it will be worthwhile to look into the issue. One possibility is that the gauge-fixing sensitiveness might have been caused by incomplete gauge-fixing: for example the presence of the trace mode of the fluctuation metric \cite{Park:2015xoa}. As a matter of fact, we believe that invesitgation of potential troubles by the trace mode should also be worth in other contexts such as massive gravity.

\newpage

\appendix

\renewcommand{\theequation}{A.\arabic{equation}}
\setcounter{equation}{0}
\section{Jet bundle and derivation of \rf{conn}}

In this appendix the derivation of \rf{conn} will be presented for self-containedness of the present work; we will mostly follow the mathematical account in \cite{Fatibene}.

Consider a principal bundle ${\cal P}=(P,M,\pi,G)$ where $P$ is the total space, $M$ is the base, $\pi$ is the projection and $G$ is the standard fiber. Let us denote by $\G_x({\cal P})$ the set of all local sections of ${\cal P}$ whose domain contains $x\in M$. Let us introduce an equivalence class by a criterion of having the same first order Taylor expansion at $x\in M$. The equivalence class associated with the section $\r$ of  $\G_x(\cal P)$ will be denoted by $j_x^1\r$. Let us denote by $J^1_x P$ the quotient space of all equivalence classes ($\sim$ denotes the equivalence class of $j_x^1\r$),
\bea
J^1_x P = \G_x(\cal P)/\sim
\eea 
The first order jet bundle $J^1P$ is a collection (more precisely, a disjoint union) of $J^1_x P$:
\bea
J^1P=\{J_x^1 P|x\in M\}
\eea
Higher order jet bundles $J^kP, k\geq 0$ (with $J^0 P=P$) are similarly defined.
A trivialization of a fiber bundle ${\cal P}$ with the fibered coordinates, $(x^\m,h^a)$, introduces the so-called natural coordinates of $J^1P$, $(x^\m,h^a,h_\m^a)$. In the context of a field theory, the fiber coordinate $h^a$ represents the fields and the derivative coordinates $h_\m^a$ represent the first derivative of the fields, or the metric and derivatives of the metric respectively in the present case.

Consider a vector field on the total space $P$,
\bea
\bfZ=\z^\m\pa_\m+\z^a \pa_a \label{gv}
\eea
Under the group action, the components $\z^\m,\z^a$ transform
\bea
\z^{\m'} &=&  \fr{\pa x^{\m'}}{\pa x^\n} \z^\n  \nn\\
\z^{a'}  &=&  \fr{\pa h^{a'}}{\pa x^\n}  \z^\n+  \fr{\pa h^{a'}}{\pa h^b}\z^b  \nn\\
{h}_{\m'}^{a'}&=& \fr{\pa x^{\n}}{\pa x^{\m'}}\Big(\fr{\pa h^{a'}}{\pa x^\n}+\fr{\pa h^{a'}}{\pa h^b} h_\n^b \Big) \la{compotrans}
\eea
The transformation of the connection component (denoted by $\w_\m^a$ below) can be deduced from the third equation of \rf{compotrans} as we will discuss shortly. 
We now turn to the definition of the connection through the horizontal projection.

There are several closely related ways to define a general (as opposed to principal) connection in a principal bundle. The connection $H$ may be defined as a map that assigns the horizontal space to each point in the bundle, $u\in P$:
\bea
H:u\ra H_u \subset T_uP
\eea
where $H_u$ is the subspace of the tangent space $T_uP$.
One can also view the connection as a distribution (i.e., assignment to each point $u\in P$ of the subspace of the tangent space at $u$, $T_uP$.) of ${\cal P}$. 
The image, denoted by ${\cal H}$, of the connection is a sub-bundle of the tangent bundle $T_uP$. There is a special class of connections called the principal connections with an additional condition - called equivariance - that at an arbitrary $u$
\bea
(T_uR_g)H_u=H_{u\cdot g}\;\;,\;\; g\in G
\eea
The abstract concept of connection becomes more tangible with the introduction of a connection one-form $\w(x,h)$ that is vector-valued: it takes, in the fibered coordinates $(x^\m,h^a)$,
\bea
\w(x,h)=dx^\m\otimes (\pa_\m+\w_\m^a(x,h)\pa_a)
\eea
Let us also introduce the basis of the Lie algebra (the set of left-invariant vector fields), $T_A$. The two bases $T_A$ and $\pa_a$ are related by a linear transformation through a matrix that we will denote by $T_A^a$:
\bea
T_A=T_A^a \pa_a
\eea
For the definition of the principal connection that we will shortly give, a set of right-invariant vertical vector fields, $\r_A$, are needed as well:
\bea
\r_A=R_A^a \pa_a
\eea
where $R_A^a$ is a matrix associated with the right multiplication.
One can show that the connection becomes a principal connection iff $\w_\m^i(u)$ becomes independent of the $h^a$ coordinates:
\bea
\w=dx^\m\otimes \Big[\pa_\m+\w_\m^A(x)\r_A \Big]
\eea
In a local coordinate system, this takes
\bea
\w=dx_{(\a)}^\m\otimes \Big[\pa_\m^{(\a)}+(\w^{(\a)})_\m^A(x)\r_A^{(\a)}\Big]
\eea
The transformations \rf{compotrans} imply that the connection field transforms, under a coordinate transformation,
\bea
(\w^{(\b)})_{\m'}^A=  \fr{\pa x^{\n}}{\pa x^{\m'}}\Big[Ad_B^A(\vf_{(\b\a)}) (\w^{(\a)})_\n^B
+(R^{-1})_a^A(\vf_{(\b\a)})\pa_\n \vf_{(\b\a)}^a \label{wtrans}
\Big]
\eea
where `$Ad$' denotes an adjoint representation and corresponds to $\fr{\pa h^{a'}}{\pa h^b}$ in \rf{compotrans}. Let us now consider $J^1P$ with its natural coordinates  $(x^\m,h^a,h_\m^a)$;
one can choose in each orbit $[x^\m,h^a,h_\m^a]_G$ a representative of the form
\bea
[x^\m,e^a,H_\m^A]_G
\eea
where $e$ is the identity element of $G$ and $H_\m^A\equiv (R^{-1})_a^A(h)h_\m^a$. The projection $\pi_{J^1P/G}:(x^\m,H_\m^A)\ra (x^\m)$ then leads to the bundle $J^1P/G$. Once again the transformation rules in \rf{compotrans} imply the following action of the transition function $\vf_{\b\a}$:
\bea
(H^{(\b)})_{\m'}^A=\fr{\pa x^{\n}}{\pa x^{\m'}}\Big[Ad_B^A(\vf_{(\b\a)}) (H^{(\a)})_\n^B
+(R^{-1})_a^A(\vf_{(\b\a)})\pa_\n \vf_{(\b\a)}^a \Big] \label{thtrans}
\eea
Since $\w_\m^A$ and $H_\m^A$ obey the same transformation rules under the transition functions, they can be globally identified: $J^1P/G={\cal C}P$ is established.

\renewcommand{\theequation}{B.\arabic{equation}}
 \setcounter{equation}{0}
\section{Derivation of \rf{constronn}}

The covariant derivative in \rf{mtmconstr} yields zero when it acts on terms 
 other than $\fr1{n}$. Consider for instance the first term inside the parenthesis in \rf{mtmconstr}\footnote{{The covariant derivative $\nabla_a$ is three-dimensional whereas the Lie derivative $\mathscr{L}_{\pa_{3}}$ is four-dimensional. This potentially poses a subtlety in applying \rf{liecocompr}. Conceptually speaking, the action of the Lie derivative on the 3D covariant derivative will be well-defined since the Lie derivative is based on the one-parameter family of group transformations that should have well defined action on the 3D quantities. More quantitatively, it is possible to understand the 3D covariant derivative as arising from the 4D one through projection that is implemented by $e_a^\a$ as indicated. (It should be possible to reach the same conclusion in the context of the coordinate-free analysis. To that end, one would consider a curve horizontally lifted to the total space built on the 3D spacetime. Still another coordinate-free method would be to use the projection operator defined in \cite{Gourgoulhon:2007ue}: it is denoted by $\vec{\g}^*$ and a coordinate-free version of $e_a^\a$.)}
 The relations between the 3D and 4D quantities become simplified partially due to our simply adapted 3D coordinates, $y_{3D}^m=x_{4D}^{m} (=y^m)$.
In other words, the second term of the far right-hand side of \rf{liecocompr} vanishes for a simply adapted 3D coordinate system such as the present one. The relation \rf{liecocompresent} would be modified according to that second term in general.} 
:  
\bea
 \N_a \mathscr{L}_{\pa_{3}} \g_{bc}=e_a^\a  \N_\a \mathscr{L}_{\pa_{3}} \g_{bc}
 \la{liecocompresent}
\eea
where $\pa_3$ denotes $\fr{\pa}{\pa x_3}$.
By using \rf{liecocom}, the right-hand side can be written as
\bea
 =e_a^\a   \mathscr{L}_{\pa_{3}} \N_\a \g_{bc}
\eea
This is because $\N_{[\bfX,\bfY]}=\N_{[\pa_{3},\pa_\a]}=\N_{0}=0$ due to the linearity of $\N$. On account of $\mathscr{L}_{\pa_{3}} e_a^\a=0 $ (see, e.g., below (2.23) of \cite{Poisson}), the right-hand side becomes
\bea
= \mathscr{L}_{\pa_{3}} e_a^\a  \N_\a \g_{bc} =\mathscr{L}_{\pa_{3}}   \N_a \g_{bc}=0
\eea
where the last equality follows from the 3D metric compatibility of the 3D 
covariant derivative.

\newpage


\begin{thebibliography}{99}
\bibitem{Goldberg} 

T. Goldberg,``What is a connection, and what is it good for?," Lecture note
\end{thebibliography}
\end{document}